\documentstyle[11pt,ysc,twoside]{article}
\def\edcomment#1{\iffalse\marginpar{\raggedright\sl#1\/}\else\relax\fi}
\reversemarginpar

\begin{document}
\title{Catalogue of Planetary Objects. Version 2006.0}

\author{Zakhozhay O.V.}
\affil{V.N.Karazin Kharkiv National University, 4 Svobody Sq.,
Kharkiv 61077, Ukraine}

\author{Zakhozhay V.A., Krugly Yu.N.}
\affil{V.N.Karazin Kharkiv National University, 4 Svobody Sq.,
Kharkiv 61077, Ukraine}

\begin{abstract}
The analysis of the density and brightness of big planets'
satellites, main asteroid belt objects, Kuiper belt objects and
centaurs has been carried out as well as the analysis of suspected
unseen satellites of the stars. According to the date on the first
of January, 2006 the catalogue of planetary objects has been
compiled.
\end{abstract}


Planets are the cosmic bodies, which have mass within limits
$(10^{-11}-10^{-10})\div10^{-2} m_{\sun}$ (Alexandrov Yu.V. \&
Zakhozhay V.A. 1980; Sluta E.N. \& Voropaev S.A. 1980; Zakhozhay
V.A. 2005a). Based on this criterion, the analysis of the masses of
Solar system objects (big planets' satellites, main asteroid belt
objects, Kuiper belt objects and centaurs) has been carried out as
well as the analysis of suspected unseen satellites of the stars.
Masses of Solar system bodies were based on analyze from space
mission data, brightness and geometric albedo, based on assumption
that average density of silicate planets --- 3--5 g/cm$^{3}$, ice
--- 1 g/cm$^{3}$, and geometric albedo of Kuiper belt objects are
within the limits 0.03--0.12 (Zakhozhay O.V., Zakhozhay V.A., \&
Krugly Yu.N. 2006a;Zakhozhay O.V., Zakhozhay V.A., \& Krugly Yu.N.
2006b). The exoplanet masses based on observation parameter
$m_{p}\cdot\mbox{\it Sini}$ were analyzed.

As a result, according to the date on the first of January 2006
the catalogue of planetary objects has been compiled. It consists
of 9 tables. The first three tables consist of the planetary
bodies. Whose belonging to planets is beyond doubt. They are: big
planets (Table 1.); 20 satellites of the big planets (Table 2.),
whose masses exceed the minimal masses of the planets
--- $m_{po}= (10^{-11}-10^{-10})m_{\sun}$; 3 Main asteroid belt
objects (Table 3.), whose masses exceed $m_{po}$ (based on the
calculations according to the albedo and brightness). Estimation
of centaurs and Kuiper belt objects masses were based on absolute
brightness (which was taken from database
[http://cfa-www.harvard.edu]) and calculations of their albedo and
average density. The list of these candidates includes four tables
(depending on heliocentric distance): $a<30$ a.u. (centaurs) ---
Table 4.; $a = 40\pm10$ a.u. (general Kuiper belt) --- Table 5.;
$50 < a < 80$ a.u. (near region of scattered-disk Kuiper belt)
--- Table 6.; $a < 80$ a.u. (distant region of scattered-disk
Kuiper belt) --- Table 7. The most probable discovered exoplanets
are itemized in the Table 8. Table 9. represents a list of the
candidates which existence wasn't confirmed. The exoplanets data
was taken from works (Zakhozhay V.A. 2005b; Zakhozhay V.A. 2001)
and database [http://www.obspm.fr/planets].

\begin{table}[htb]
\begin{center}
\caption{Big planets}
\begin{tabular}
{|c|c|c|c|c|c|} \hline N& Name& $m_{p}$, g& $D$, km& $\rho$,
g/cm$^{3}$&
$a$, a.u. \\
\hline 1.1& Mercury& 3.30 $\cdot $ 10$^{26}$& 4879 \quad & 5.43&
0.387 \\
\hline 1.2& Venus& 4.87 $\cdot $ 10$^{27}$& 12104& 5.24&
0.723 \\
\hline 1.3& Earth& 5.97 $\cdot $ 10$^{27}$& 12756& 5.52&
1.000 \\
\hline 1.4& Mars& 6.42 $\cdot $ 10$^{26}$& 6794& 3.94&
1.524 \\
\hline 1.5& Jupiter& 1.90 $\cdot $ 10$^{30}$& 142984& 1.33&
5.204 \\
\hline 1.6& Saturn& 5.69 $\cdot $ 10$^{29}$& 120536& 0.70&
9.584 \\
\hline 1.7& Uranus& 8.66 $\cdot $ 10$^{28}$& 51118& 1.30&
19.187 \\
\hline 1.8& Neptune& 1.03 $\cdot $ 10$^{29}$& 49528& 1.76&
30.021 \\
\hline 1.9& Pluto& 1.50 $\cdot $ 10$^{25}$& 2390& 1.1&
39.231 \\
\hline
\end{tabular}
\label{tab1}
\end{center}
\end{table}

\begin{table}[htb]
\begin{center}
\caption{Satellites}
\begin{tabular}
{|c|c|c|c|c|c|} \hline N& Name& $m_{p}$, g& $D$, km& $\rho$,
g/cm$^{3}$&
$a$, a.u. \\
\hline 2.1& Moon& 7.35 $\cdot $ 10$^{25}$& 3475& 3.34&
2.570 $ \cdot $ 10$^{ - 3}$ \\
\hline 2.2& Io& 8.92 $\cdot $ 10$^{25}$& 3640& 3.55&
2.821 $ \cdot $ 10$^{ - 3}$ \\
\hline 2.3& Europa& 4.85 $\cdot $ 10$^{25}$& 3130& 3.04&
4.485 $ \cdot $ 10$^{ - 3}$ \\
\hline 2.4& Ganymede& 1.49 $\cdot $ 10$^{26}$& 5268& 1.93&
7.153 $ \cdot $ 10$^{ - 3}$ \\
\hline 2.5& Callisto& 1.08 $\cdot $ 10$^{26}$& 4806& 1.83&
12.587 $ \cdot $ 10$^{ - 3}$ \\
\hline 2.6& Mimas& 3.80 $\cdot $ 10$^{22}$& 390& 1.2&
1.240 $ \cdot $ 10$^{ - 3}$ \\
\hline 2.7& Enceladus& 7.00 $\cdot $ 10$^{22}$& 500& 1.1&
1.591 $ \cdot $ 10$^{ - 3}$ \\
\hline 2.8& Tethys& 7.60 $\cdot $ 10$^{23}$& 1060& 1.2&
1.970 $ \cdot $ 10$^{ - 3}$ \\
\hline 2.9& Dione& 1.05 $\cdot $ 10$^{23}$& 1120& 1.4&
2.523 $ \cdot $ 10$^{ - 3}$ \\
\hline 2.10& Rhea& 2.49 $\cdot $ 10$^{24}$& 1530& 1.3&
3.523 $ \cdot $ 10$^{ - 3}$ \\
\hline 2.11& Titan& 1.35 $\cdot $ 10$^{26}$& 5150& 1.9&
8.167 $ \cdot $ 10$^{ - 3}$ \\
\hline 2.12& Iapetus& 1.75 $\cdot $ 10$^{24}$& 1440& 1.2&
23.806 $ \cdot $ 10$^{ - 3}$ \\
\hline 2.13& Miranda& 7.00 $\cdot $ 10$^{22}$& 470& 1.3&
0.865 $ \cdot $ 10$^{ - 3}$ \\
\hline 2.14& Ariel& 1.36 $\cdot $ 10$^{24}$& 1160& 1.6&
1.277 $ \cdot $ 10$^{ - 3}$ \\
\hline 2.15& Umbriel& 1.20 $\cdot $ 10$^{24}$& 1190& 1.4&
1.780 $ \cdot $ 10$^{ - 3}$ \\
\hline
\end{tabular}
\end{center}
\label{tab2}
\end{table}

\begin{table}[htb]
\begin{center}
\begin{tabular}
{|c|c|c|c|c|c|} \hline N& Name& $m_{p}$, g& $D$, km& $\rho$,
g/cm$^{3}$&
$a$, a.u. \\
\hline 2.16& Titania& 3.53 $\cdot $ 10$^{24}$& 1580& 1.6&
2.914 $ \cdot $ 10$^{ - 3}$ \\
\hline 2.17& Oberon& 3.00 $\cdot $ 10$^{24}$& 1530& 1.5&
3.901 $ \cdot $ 10$^{ - 3}$ \\
\hline 2.18& Triton& 2.15 $\cdot $ 10$^{25}$& 2700& 2.1&
2.371 $ \cdot $ 10$^{ - 3}$ \\
\hline 2.19& Nereid& 2.00 $\cdot $ 10$^{22}$& 340& 1.0&
36.855 $ \cdot $ 10$^{ - 3}$ \\
\hline 2.20& Charon& 1.80 $\cdot $ 10$^{24}$& 1200& 2.1&
0.131 $ \cdot $ 10$^{ - 3}$ \\
\hline
\end{tabular}
\end{center}
\end{table}

\begin{table}[htb]
\begin{center}
\caption{Asteroids}
\begin{tabular}
{|c|c|c|c|c|c|} \hline N& Name& $m_{p}$, g& $D$, km& $\rho
,$g/cm$^{3}$&
$a$, a.u. \\
\hline 3.1& Ceres& 9.47 $\cdot $ 10$^{23}$& 960& 2.0&
2.767 \\
\hline 3.2& Pallas& 2.15 $\cdot $ 10$^{23}$& 550& 2.5&
2.773 \\
\hline 3.3& Vesta& 2.67 $\cdot $ 10$^{23}$& 530& 3.4&
2.361 \\
\hline
\end{tabular}
\end{center}
\label{tab3}
\end{table}

\begin{table}[htb]
\caption{Centaurs}
\begin{tabular}
{|p{19pt}|p{60pt}|p{27pt}|p{30pt}|p{0pt}|p{19pt}|p{60pt}|p{26pt}|p{30pt}|}
\cline{1-4} \cline{6-9} N& Name& $H$& $a$, a.u.& ~& N& Name& $H$&
$a$, a.u. \\
\cline{1-4} \cline{6-9} 4.1& 1995 SN55& 6.0& 23.564& ~& 4.4& 2002
GZ32& 6.8&
23.212 \\
\cline{1-4} \cline{6-9} 4.2& 1997 CU26 $^{1}$ & 6.4& 15.868& ~&
4.5& 1992 AD$^{ 3}$ & 7.0&
20.431 \\
\cline{1-4} \cline{6-9} 4.3& 1977 UB $^{2}$ & 6.5& 13.685& ~& ~&
~& ~&
~ \\
\cline{1-4} \cline{6-9}
\end{tabular}
\label{tab4}
\end{table}

\begin{table}[htbp]
\caption{General Kuiper belt}
\begin{tabular}
{|p{20pt}|p{68pt}|p{18pt}|p{30pt}|p{0pt}|p{22pt}|p{70pt}|p{14pt}|p{30pt}|}
\cline{1-4} \cline{6-9} N& Name& $H$& $a$, a.u.& ~& N& Name& $H$&
$a$, a.u. \\
\cline{1-4} \cline{6-9} 5.1& 2005 FY9& $-0.4$& 45.706& & 5.261&
2005 JO179& 6.7&
43.633 \\
\cline{1-4} \cline{6-9} 5.2& 2003 EL61& 0.1& 43.338& & 5.262& 2005
GV210& 6.7&
45.015 \\
\cline{1-4} \cline{6-9} 5.3& 2004 DW $^{4}$ & 2.3& 39.386& &
5.263& 2005 GE187& 6.7&
39.6 \\
\cline{1-4} \cline{6-9} 5.4& 2002 LM60 $^{5}$ & 2.6& 43.548& &
5.264& 2005 GB187& 6.7&
39.795 \\
\cline{1-4} \cline{6-9} 5.5& 2001 KX76 $^{6}$ & 3.2& 39.623& &
5.265& 2004 PF112& 6.7&
44.925 \\
\cline{1-4} \cline{6-9} 5.6& 2002 TX300& 3.3& 43.088& & 5.266&
2004 PE112& 6.7&
45.903 \\
\cline{1-4} \cline{6-9} 5.7& 2002 AW197& 3.3& 47.37& & 5.267& 2004
PU107& 6.7&
44.997 \\
\cline{1-4} \cline{6-9} 5.8& 2002 UX25& 3.6& 42.526& & 5.268& 2003
UK293& 6.7&
43.92 \\
\cline{1-4} \cline{6-9} 5.9& 2000 WR106 $^{7}$ & 3.7& 42.954& &
5.269& 2003 UE292& 6.7&
43.469 \\
\cline{1-4} \cline{6-9} 5.10& 2002 MS4& 3.8& 41.864& & 5.270& 2003
QT91& 6.7&
44.093 \\
\cline{1-4} \cline{6-9} 5.11& 2003 AZ84& 3.9& 39.508& & 5.271&
2003 QL91& 6.7&
42.946 \\
\cline{1-4} \cline{6-9} 5.12& 2004 GV9& 4.0& 42.252& & 5.272& 2003
QH91& 6.7&
39.187 \\
\cline{1-4} \cline{6-9} 5.13& 2003 OP32& 4.1& 43.183& & 5.273&
2003 QZ90& 6.7&
43.577 \\
\cline{1-4} \cline{6-9} 5.14& 2003 VS2& 4.2& 39.289& & 5.274& 2003
FC128& 6.7&
35.127 \\
\cline{1-4} \cline{6-9} 5.15& 2003 QW90& 4.4& 43.621& & 5.275&
2003 FM127& 6.7&
43.304 \\
\cline{1-4} \cline{6-9} 5.16& 2004 SB60& 4.4& 41.92& & 5.276& 2002
VF131& 6.7&
44.025 \\
\cline{1-4} \cline{6-9} 5.17& 2002 KX14& 4.4& 39& & 5.277& 2002
VA131& 6.7&
42.111 \\
\cline{1-4} \cline{6-9} 5.18& 2004 TY364& 4.5& 38.708& & 5.278&
2002 PT170& 6.7&
46.154 \\
\cline{1-4} \cline{6-9} 5.19& 1996 TO66& 4.5& 43.161& & 5.279&
2001 QG298& 6.7&
39.194 \\
\cline{1-4} \cline{6-9} 5.20& 2004 PR107& 4.6& 45.75& & 5.280&
2001 QQ297& 6.7&
44.24 \\
\cline{1-4} \cline{6-9} 5.21& 2002 KW14& 4.6& 46.959& & 5.281&
2001 KU76& 6.7&
45.347 \\
\cline{1-4} \cline{6-9} 5.22& 2001 QF298& 4.7& 39.268& & 5.282&
2001 KO76& 6.7&
43.915 \\
\cline{1-4} \cline{6-9}
\end{tabular}
\end{table}

\begin{table}[htbp]
\begin{tabular}
{|p{20pt}|p{68pt}|p{18pt}|p{30pt}|p{0pt}|p{22pt}|p{70pt}|p{14pt}|p{30pt}|}
\cline{1-4} \cline{6-9} N& Name& $H$& $a$, a.u.& ~& N& Name& $H$&
$a$, a.u. \\
\cline{1-4} \cline{6-9} 5.23& 2000 EB173 $^{8}$ & 4.7& 39.77& &
5.283& 2000 CP104& 6.7&
44.47 \\
\cline{1-4} \cline{6-9} 5.24& 1995 SM55& 4.8& 41.65& & 5.284& 1999
RN215& 6.7&
43.01 \\
\cline{1-4} \cline{6-9} 5.25& 1999 TC36& 4.9& 39.242& & 5.285&
1999 RE215& 6.7&
44.828 \\
\cline{1-4} \cline{6-9} 5.26& 1998 WH24 $^{9}$ & 4.9& 45.599& &
5.286& 1998 WY24& 6.7&
43.121 \\
\cline{1-4} \cline{6-9} 5.27& 2002 XV93& 5.0& 39.263& & 5.287&
1998 FS144& 6.7&
42.007 \\
\cline{1-4} \cline{6-9} 5.28& 2000 CN105& 5.0& 44.713& & 5.288&
1996 RR20& 6.7&
39.431 \\
\cline{1-4} \cline{6-9} 5.29& 2003 FY128& 5.0& 49.84& & 5.289&
2001 KE76& 6.7&
43.226 \\
\cline{1-4} \cline{6-9} 5.30& 2002 CY248& 5.1& 46.243& & 5.290&
1999 HX11 $^{11}$ & 6.7&
39.21 \\
\cline{1-4} \cline{6-9} 5.31& 1999 CD158& 5.1& 43.755& & 5.291&
2005 JH177& 6.8&
44.002 \\
\cline{1-4} \cline{6-9} 5.32& 2002 WC19& 5.1& 47.73& & 5.292& 2005
GW186& 6.8&
43.213 \\
\cline{1-4} \cline{6-9} 5.33& 1997 CS29& 5.1& 43.943& & 5.293&
2005 EB318& 6.8&
44.522 \\
\cline{1-4} \cline{6-9} 5.34& 2003 UZ117& 5.2& 44.054& & 5.294&
2004 VA76& 6.8&
43.946 \\
\cline{1-4} \cline{6-9} 5.35& 2003 QM91& 5.3& 45.247& & 5.295&
2004 EU95& 6.8&
44.454 \\
\cline{1-4} \cline{6-9} 5.36& 2003 QA91& 5.3& 44.269& & 5.296&
2003 UP292& 6.8&
39.276 \\
\cline{1-4} \cline{6-9} 5.37& 2002 PJ149& 5.3& 45.361& & 5.297&
2003 QZ111& 6.8&
42.895 \\
\cline{1-4} \cline{6-9} 5.38& 2002 GV31& 5.3& 43.253& & 5.298&
2003 QD91& 6.8&
43.158 \\
\cline{1-4} \cline{6-9} 5.39& 2001 KA77& 5.3& 47.645& & 5.299&
2003 QX90& 6.8&
43.747 \\
\cline{1-4} \cline{6-9} 5.40& 2002 VE95& 5.3& 39.154& & 5.300&
2003 FF128& 6.8&
39.844 \\
\cline{1-4} \cline{6-9} 5.41& 2001 UQ18& 5.4& 44.083& & 5.301&
2003 FH127& 6.8&
44.817 \\
\cline{1-4} \cline{6-9} 5.42& 2001 QS322& 5.4& 43.889& & 5.302&
2002 PA171& 6.8&
44.261 \\
\cline{1-4} \cline{6-9} 5.43& 2001 QY297& 5.4& 43.855& & 5.303&
2002 GY32& 6.8&
39.736 \\
\cline{1-4} \cline{6-9} 5.44& 2001 QS297& 5.4& 43.797& & 5.304&
2002 FV6& 6.8&
47.228 \\
\cline{1-4} \cline{6-9} 5.45& 2000 CO105& 5.4& 47.097& & 5.305&
2002 CB225& 6.8&
44.307 \\
\cline{1-4} \cline{6-9} 5.46& 2005 PR21& 5.5& 44.857& & 5.306&
2001 XB255& 6.8&
44.717 \\
\cline{1-4} \cline{6-9} 5.47& 2002 XH91& 5.5& 43.924& & 5.307&
2001 RV143& 6.8&
43.085 \\
\cline{1-4} \cline{6-9} 5.48& 2002 GJ32& 5.5& 44.521& & 5.308&
2001 QE298& 6.8&
43.543 \\
\cline{1-4} \cline{6-9} 5.49& 2001 YH140& 5.5& 42.384& & 5.309&
2001 OS108& 6.8&
46.453 \\
\cline{1-4} \cline{6-9} 5.50& 2001 XD255& 5.5& 39.504& & 5.310&
2001 KQ76& 6.8&
42.842 \\
\cline{1-4} \cline{6-9} 5.51& 2001 QT297& 5.5& 43.965& & 5.311&
2001 FO185& 6.8&
46.827 \\
\cline{1-4} \cline{6-9} 5.52& 2002 XW93& 5.5& 37.422& & 5.312&
2000 YB2& 6.8&
38.632 \\
\cline{1-4} \cline{6-9} 5.53& 2004 PT107& 5.6& 40.538& & 5.313&
2000 CE105& 6.8&
43.909 \\
\cline{1-4} \cline{6-9} 5.54& 2002 VU130& 5.6& 39.065& & 5.314&
1999 RX214& 6.8&
45.913 \\
\cline{1-4} \cline{6-9} 5.55& 2002 VT130& 5.6& 42.314& & 5.315&
1999 OJ4& 6.8&
38.01 \\
\cline{1-4} \cline{6-9} 5.56& 2002 PM149& 5.6& 44.144& & 5.316&
1999 OX3& 6.8&
32.136 \\
\cline{1-4} \cline{6-9} 5.57& 2002 PQ145& 5.6& 43.773& & 5.317&
2001 KJ76& 6.8&
43.911 \\
\cline{1-4} \cline{6-9} 5.58& 2002 GD32& 5.6& 44.737& & 5.318&
1999 OY3& 6.8&
43.878 \\
\cline{1-4} \cline{6-9} 5.59& 2001 XR254& 5.6& 42.988& & 5.319&
1998 KR65& 6.8&
43.496 \\
\cline{1-4} \cline{6-9} 5.60& 2001 QD298& 5.6& 42.59& & 5.320&
2001 FM185& 6.8&
38.784 \\
\cline{1-4} \cline{6-9} 5.61& 2005 JA175& 5.7& 42.728& & 5.321&
1996 TP66& 6.8&
39.201 \\
\cline{1-4} \cline{6-9} 5.62& 2003 UB292& 5.7& 46.904& & 5.322&
2005 PO21& 6.9&
43.291 \\
\cline{1-4} \cline{6-9} 5.63& 2003 GH55& 5.7& 44.475& & 5.323&
2005 GX206& 6.9&
39.622 \\
\cline{1-4} \cline{6-9} 5.64& 2002 GF32& 5.7& 39.517& & 5.324&
2005 GD187& 6.9&
44.555 \\
\cline{1-4} \cline{6-9} 5.65& 2001 CZ31& 5.7& 45.41& & 5.325& 2005
GC187& 6.9&
44.568 \\
\cline{1-4} \cline{6-9} 5.66& 2002 VR128& 5.7& 39.272& & 5.326&
2005 EN302& 6.9&
45.083 \\
\cline{1-4} \cline{6-9} 5.67& 2005 JP179& 5.8& 43.419& & 5.327&
2004 VM78& 6.9&
44.359 \\
\cline{1-4} \cline{6-9} 5.68& 2003 WU188& 5.8& 44.072& & 5.328&
2004 VZ75& 6.9&
43.673 \\
\cline{1-4} \cline{6-9} 5.69& 2003 QA112& 5.8& 42.648& & 5.329&
2004 VY75& 6.9&
45.532 \\
\cline{1-4} \cline{6-9}
\end{tabular}
\end{table}

\begin{table}[htbp]
\begin{tabular}
{|p{20pt}|p{68pt}|p{18pt}|p{30pt}|p{0pt}|p{22pt}|p{70pt}|p{14pt}|p{30pt}|}
\cline{1-4} \cline{6-9} N& Name& $H$& $a$, a.u.& ~& N& Name& $H$&
$a$, a.u. \\
\cline{1-4} \cline{6-9} 5.70& 2002 GH32& 5.8& 42.251& & 5.330&
2004 PX107& 6.9&
45.169 \\
\cline{1-4} \cline{6-9} 5.71& 2001 RU143& 5.8& 39.297& & 5.331&
2004 ES95& 6.9&
45.011 \\
\cline{1-4} \cline{6-9} 5.72& 2001 QO297& 5.8& 42.771& & 5.332&
2004 EP95& 6.9&
43.409 \\
\cline{1-4} \cline{6-9} 5.73& 2001 KD77& 5.8& 39.766& & 5.333&
2003 UC292& 6.9&
45.123 \\
\cline{1-4} \cline{6-9} 5.74& 2000 AF255& 5.8& 48.786& & 5.334&
2003 QB112& 6.9&
43.474 \\
\cline{1-4} \cline{6-9} 5.75& 1999 KR16& 5.8& 49.241& & 5.335&
2003 QV90& 6.9&
43.668 \\
\cline{1-4} \cline{6-9} 5.76& 1998 SN165& 5.8& 37.764& & 5.336&
2003 LD9& 6.9&
47.591 \\
\cline{1-4} \cline{6-9} 5.77& 1998 SM165& 5.8& 47.446& & 5.337&
2003 LD7& 6.9&
42.9 \\
\cline{1-4} \cline{6-9} 5.78& 2005 JR179& 5.9& 45.488& & 5.338&
2003 GM53& 6.9&
44.454 \\
\cline{1-4} \cline{6-9} 5.79& 2004 VT75& 5.9& 39.991& & 5.339&
2003 FB128& 6.9&
39.872 \\
\cline{1-4} \cline{6-9} 5.80& 2004 PV107& 5.9& 44.74& & 5.340&
2002 XF91& 6.9&
44.615 \\
\cline{1-4} \cline{6-9} 5.81& 2004 DH64& 5.9& 46.777& & 5.341&
2002 VW130& 6.9&
45.966 \\
\cline{1-4} \cline{6-9} 5.82& 2003 WQ188& 5.9& 46.338& & 5.342&
2002 GW32& 6.9&
35.195 \\
\cline{1-4} \cline{6-9} 5.83& 2003 FE128& 5.9& 48.342& & 5.343&
2002 GU32& 6.9&
39.538 \\
\cline{1-4} \cline{6-9} 5.84& 2001 UP18& 5.9& 47.57& & 5.344& 2002
FX6& 6.9&
45.113 \\
\cline{1-4} \cline{6-9} 5.85& 2001 QJ298& 5.9& 44.014& & 5.345&
2002 CW224& 6.9&
39.235 \\
\cline{1-4} \cline{6-9} 5.86& 2000 CJ105& 5.9& 44.369& & 5.346&
2001 FK193& 6.9&
44.346 \\
\cline{1-4} \cline{6-9} 5.87& 1999 CL119& 5.9& 47.013& & 5.347&
2001 FQ185& 6.9&
48.087 \\
\cline{1-4} \cline{6-9} 5.88& 1999 RZ253& 5.9& 43.721& & 5.348&
2001 DD106& 6.9&
44.273 \\
\cline{1-4} \cline{6-9} 5.89& 2002 XE91& 6.0& 45.59& & 5.349& 2000
YA2& 6.9&
43.968 \\
\cline{1-4} \cline{6-9} 5.90& 2002 PD149& 6.0& 42.83& & 5.350&
2000 YX1& 6.9&
45.429 \\
\cline{1-4} \cline{6-9} 5.91& 2002 CX224& 6.0& 46.034& & 5.351&
2000 PD30& 6.9&
46.474 \\
\cline{1-4} \cline{6-9} 5.92& 2002 CP154& 6.0& 43.413& & 5.352&
2000 CO114& 6.9&
46.143 \\
\cline{1-4} \cline{6-9} 5.93& 2001 RZ143& 6.0& 43.868& & 5.353&
2000 CF105& 6.9&
43.95 \\
\cline{1-4} \cline{6-9} 5.94& 2001 QU297& 6.0& 43.706& & 5.354&
2000 CN104& 6.9&
42.927 \\
\cline{1-4} \cline{6-9} 5.95& 2000 ON67& 6.0& 42.878& & 5.355&
1999 RC215& 6.9&
43.853 \\
\cline{1-4} \cline{6-9} 5.96& 2000 OK67& 6.0& 46.398& & 5.356&
1999 GS46& 6.9&
44.82 \\
\cline{1-4} \cline{6-9} 5.97& 2000 OJ67& 6.0& 42.758& & 5.357&
1999 OF4& 6.9&
44.914 \\
\cline{1-4} \cline{6-9} 5.98& 2000 KK4& 6.0& 41.48& & 5.358& 1999
CL158& 6.9&
41.571 \\
\cline{1-4} \cline{6-9} 5.99& 2000 GN171& 6.0& 39.726& & 5.359&
1994 JQ1& 6.9&
44.379 \\
\cline{1-4} \cline{6-9} 5.100& 2005 EF298& 6.1& 44.991& & 5.360&
1993 SC& 6.9&
39.367 \\
\cline{1-4} \cline{6-9} 5.101& 2004 PY107& 6.1& 44.231& & 5.361&
2005 PK21& 7.0&
44.329 \\
\cline{1-4} \cline{6-9} 5.102& 2004 OL12& 6.1& 43.632& & 5.362&
2005 GZ186& 7.0&
43.83 \\
\cline{1-4} \cline{6-9} 5.103& 2004 EO95& 6.1& 43.746& & 5.363&
2005 EX318& 7.0&
39.611 \\
\cline{1-4} \cline{6-9} 5.104& 2003 UJ292& 6.1& 43.779& & 5.364&
2004 VS75& 7.0&
43.473 \\
\cline{1-4} \cline{6-9} 5.105& 2003 UA292& 6.1& 46.767& & 5.365&
2004 UT10& 7.0&
47.85 \\
\cline{1-4} \cline{6-9} 5.106& 2003 QW111& 6.1& 43.591& & 5.366&
2004 DL64& 7.0&
45.353 \\
\cline{1-4} \cline{6-9} 5.107& 2002 VB131& 6.1& 42.925& & 5.367&
2003 UT292& 7.0&
39.138 \\
\cline{1-4} \cline{6-9} 5.108& 2002 CV154& 6.1& 45.665& & 5.368&
2003 UQ292& 7.0&
39.514 \\
\cline{1-4} \cline{6-9} 5.109& 2001 RX143& 6.1& 39.236& & 5.369&
2003 UN292& 7.0&
44.712 \\
\cline{1-4} \cline{6-9} 5.110& 2001 QC298& 6.1& 46.163& & 5.370&
2003 UM292& 7.0&
46.075 \\
\cline{1-4} \cline{6-9} 5.111& 2001 KY76& 6.1& 39.5& & 5.371& 2003
UX291& 7.0&
46.5 \\
\cline{1-4} \cline{6-9} 5.112& 2001 HZ58& 6.1& 42.999& & 5.372&
2003 QU91& 7.0&
43.634 \\
\cline{1-4} \cline{6-9} 5.113& 2000 WT169& 6.1& 44.773& & 5.373&
2003 QG91& 7.0&
43.881 \\
\cline{1-4} \cline{6-9} 5.114& 2000 QE226& 6.1& 43.908& & 5.374&
2003 QC91& 7.0&
43.453 \\
\cline{1-4} \cline{6-9} 5.115& 1999 XY143& 6.1& 42.915& & 5.375&
2003 FD128& 7.0&
39.4 \\
\cline{1-4} \cline{6-9} 5.116& 1998 WW31& 6.1& 44.5& & 5.376& 2002
VZ94& 7.0&
45.743 \\
\cline{1-4} \cline{6-9} 5.117& 2002 PA149& 6.1& 43.506& & 5.377&
2002 PN149& 7.0&
42.254 \\
\cline{1-4} \cline{6-9}
\end{tabular}
\label{tab5}
\end{table}

\begin{table}[htbp]
\begin{tabular}
{|p{22pt}|p{70pt}|p{14pt}|p{30pt}|p{0pt}|p{22pt}|p{70pt}|p{14pt}|p{30pt}|}
\cline{1-4} \cline{6-9} N& Name& $H$& $a$, a.u.& ~& N& Name& $H$&
$a$, a.u. \\
\cline{1-4} \cline{6-9} 5.118& 1999 DF9& 6.1& 46.679& & 5.378&
2002 PK149& 7.0&
37.113 \\
\cline{1-4} \cline{6-9} 5.119& 2005 PU21& 6.2& 44.278& & 5.379&
2002 GE32& 7.0&
39.539 \\
\cline{1-4} \cline{6-9} 5.120& 2005 PQ21& 6.2& 43.754& & 5.380&
2002 GX31& 7.0&
43.301 \\
\cline{1-4} \cline{6-9} 5.121& 2005 PL21& 6.2& 44.798& & 5.381&
2002 GW31& 7.0&
39.508 \\
\cline{1-4} \cline{6-9} 5.122& 2005 EO302& 6.2& 45.724& & 5.382&
2002 CR154& 7.0&
42.627 \\
\cline{1-4} \cline{6-9} 5.123& 2004 PA108& 6.2& 43.588& & 5.383&
2001 XV254& 7.0&
45.658 \\
\cline{1-4} \cline{6-9} 5.124& 2003 QA92& 6.2& 38.119& & 5.384&
2001 QF331& 7.0&
39.48 \\
\cline{1-4} \cline{6-9} 5.125& 2003 QR91& 6.2& 46.63& & 5.385&
2001 KQ77& 7.0&
39.723 \\
\cline{1-4} \cline{6-9} 5.126& 2003 QB91& 6.2& 39.133& & 5.386&
2001 KT76& 7.0&
45.438 \\
\cline{1-4} \cline{6-9} 5.127& 2003 LC7& 6.2& 45.14& & 5.387& 2001
FL185& 7.0&
44.359 \\
\cline{1-4} \cline{6-9} 5.128& 2003 LB7& 6.2& 45.967& & 5.388&
2000 YF2& 7.0&
45.434 \\
\cline{1-4} \cline{6-9} 5.129& 2003 LA7& 6.2& 46.563& & 5.389&
2000 YW1& 7.0&
42.92 \\
\cline{1-4} \cline{6-9} 5.130& 2002 PE155& 6.2& 43.05& & 5.390&
2000 WN183& 7.0&
44.098 \\
\cline{1-4} \cline{6-9} 5.131& 2002 FX36& 6.2& 44.381& & 5.391&
2000 QO252& 7.0&
45.556 \\
\cline{1-4} \cline{6-9} 5.132& 2001 QX297& 6.2& 44.064& & 5.392&
2000 QM251& 7.0&
44.532 \\
\cline{1-4} \cline{6-9} 5.133& 2001 QR297& 6.2& 44.323& & 5.393&
1999 OE4& 7.0&
45.287 \\
\cline{1-4} \cline{6-9} 5.134& 2000 CK105& 6.2& 39.499& & 5.394&
1999 HH12& 7.0&
43.86 \\
\cline{1-4} \cline{6-9} 5.135& 2000 CL104& 6.2& 44.617& & 5.395&
1999 CQ133& 7.0&
41.388 \\
\cline{1-4} \cline{6-9} 5.136& 1996 TC68& 6.2& 41.97& & 5.396&
1999 CX131& 7.0&
42.416 \\
\cline{1-4} \cline{6-9} 5.137& 1996 TK66& 6.2& 42.547& & 5.397&
1999 CB119& 7.0&
47.238 \\
\cline{1-4} \cline{6-9} 5.138& 2005 PN21& 6.3& 45.739& & 5.398&
1998 UU43& 7.0&
36.34 \\
\cline{1-4} \cline{6-9} 5.139& 2005 JZ174& 6.3& 45.003& & 5.399&
1996 RQ20& 7.0&
43.68 \\
\cline{1-4} \cline{6-9} 5.140& 2005 EC318& 6.3& 44.222& & 5.400&
1995 GJ& 7.0&
42.907 \\
\cline{1-4} \cline{6-9} 5.141& 2005 EO304& 6.3& 45.197& & 5.401&
1994 TG2& 7.0&
42.448 \\
\cline{1-4} \cline{6-9} 5.142& 2004 PB108& 6.3& 44.9& & 5.402&
1994 TH& 7.0&
40.94 \\
\cline{1-4} \cline{6-9} 5.143& 2004 PW107& 6.3& 43.618& & 5.403&
1994 TG& 7.0&
42.254 \\
\cline{1-4} \cline{6-9} 5.144& 2004 ER95& 6.3& 44.48& & 5.404&
1993 FW& 7.0&
44.035 \\
\cline{1-4} \cline{6-9} 5.145& 2003 UW291& 6.3& 44.614& & 5.405&
2004 EW95& 7.0&
39.724 \\
\cline{1-4} \cline{6-9} 5.146& 2003 QX111& 6.3& 39.32& & 5.406&
2003 UR292& 7.0&
32.219 \\
\cline{1-4} \cline{6-9} 5.147& 2003 QJ91& 6.3& 43.506& & 5.407&
2001 KP77& 7.0&
43.978 \\
\cline{1-4} \cline{6-9} 5.148& 2003 QY90& 6.3& 42.687& & 5.408&
2001 KN77& 7.0&
39.356 \\
\cline{1-4} \cline{6-9} 5.149& 2003 QT90& 6.3& 45.806& & 5.409&
1994 VK8& 7.0&
42.696 \\
\cline{1-4} \cline{6-9} 5.150& 2003 KO20& 6.3& 45.663& & 5.410&
2005 TN74& 7.1&
30.009 \\
\cline{1-4} \cline{6-9} 5.151& 2002 VS130& 6.3& 44.878& & 5.411&
2005 EO297& 7.1&
43.848 \\
\cline{1-4} \cline{6-9} 5.152& 2002 VE130& 6.3& 44.417& & 5.412&
2004 VX75& 7.1&
43.847 \\
\cline{1-4} \cline{6-9} 5.153& 2002 PN147& 6.3& 44.443& & 5.413&
2004 TB358& 7.1&
43.811 \\
\cline{1-4} \cline{6-9} 5.154& 2002 GN32& 6.3& 39.362& & 5.414&
2004 PA112& 7.1&
38.838 \\
\cline{1-4} \cline{6-9} 5.155& 2001 XU254& 6.3& 43.589& & 5.415&
2004 PZ111& 7.1&
43.621 \\
\cline{1-4} \cline{6-9} 5.156& 2001 QQ322& 6.3& 43.864& & 5.416&
2004 DL71& 7.1&
44.244 \\
\cline{1-4} \cline{6-9} 5.157& 2001 QB298& 6.3& 42.554& & 5.417&
2003 WV188& 7.1&
45.92 \\
\cline{1-4} \cline{6-9} 5.158& 2001 QZ297& 6.3& 44.293& & 5.418&
2003 UV292& 7.1&
39.187 \\
\cline{1-4} \cline{6-9} 5.159& 2001 QV297& 6.3& 43.976& & 5.419&
2003 UG292& 7.1&
44.113 \\
\cline{1-4} \cline{6-9} 5.160& 2001 DC106& 6.3& 43.622& & 5.420&
2003 UV291& 7.1&
44.259 \\
\cline{1-4} \cline{6-9} 5.161& 2000 QK252& 6.3& 45.147& & 5.421&
2003 UU291& 7.1&
44.161 \\
\cline{1-4} \cline{6-9} 5.162& 2000 QL251& 6.3& 47.58& & 5.422&
2003 US291& 7.1&
44.922 \\
\cline{1-4} \cline{6-9} 5.163& 2000 OU69& 6.3& 43.203& & 5.423&
2003 QB92& 7.1&
34.828 \\
\cline{1-4} \cline{6-9} 5.164& 2000 OH67& 6.3& 44.115& & 5.424&
2003 FJ127& 7.1&
44.433 \\
\cline{1-4} \cline{6-9}
\end{tabular}
\end{table}

\begin{table}[htbp]
\begin{tabular}
{|p{22pt}|p{70pt}|p{14pt}|p{30pt}|p{0pt}|p{22pt}|p{70pt}|p{14pt}|p{30pt}|}
\cline{1-4} \cline{6-9} N& Name& $H$& $a$, a.u.& ~& N& Name& $H$&
$a$, a.u. \\
\cline{1-4} \cline{6-9} 5.165& 2001 KK76& 6.3& 42.7& & 5.425& 2002
VF130& 7.1&
45.649 \\
\cline{1-4} \cline{6-9} 5.166& 2000 CM105& 6.3& 42.299& & 5.426&
2002 VD130& 7.1&
47.587 \\
\cline{1-4} \cline{6-9} 5.167& 2000 CH105& 6.3& 44.576& & 5.427&
2002 PU170& 7.1&
47.621 \\
\cline{1-4} \cline{6-9} 5.168& 2005 JQ179& 6.4& 43.784& & 5.428&
2002 PQ149& 7.1&
44.339 \\
\cline{1-4} \cline{6-9} 5.169& 2005 EW318& 6.4& 46.604& & 5.429&
2002 PP149& 7.1&
40.843 \\
\cline{1-4} \cline{6-9} 5.170& 2005 EH305& 6.4& 44.056& & 5.430&
2001 RY143& 7.1&
44.876 \\
\cline{1-4} \cline{6-9} 5.171& 2005 EX297& 6.4& 45.104& & 5.431&
2001 KM76& 7.1&
44.871 \\
\cline{1-4} \cline{6-9} 5.172& 2003 WU172& 6.4& 39.128& & 5.432&
2001 KF76& 7.1&
44.644 \\
\cline{1-4} \cline{6-9} 5.173& 2003 UT291& 6.4& 42.625& & 5.433&
2000 YV1& 7.1&
43.706 \\
\cline{1-4} \cline{6-9} 5.174& 2003 QP91& 6.4& 43.493& & 5.434&
2000 YU1& 7.1&
43.468 \\
\cline{1-4} \cline{6-9} 5.175& 2003 FL127& 6.4& 39.419& & 5.435&
2000 WV12& 7.1&
43.338 \\
\cline{1-4} \cline{6-9} 5.176& 2002 VC131& 6.4& 46.323& & 5.436&
2000 QA243& 7.1&
43.27 \\
\cline{1-4} \cline{6-9} 5.177& 2002 PW170& 6.4& 44.816& & 5.437&
1999 RT215& 7.1&
43.089 \\
\cline{1-4} \cline{6-9} 5.178& 2002 GH166& 6.4& 39.316& & 5.438&
1999 OM4& 7.1&
46.03 \\
\cline{1-4} \cline{6-9} 5.179& 2002 GK32& 6.4& 43.798& & 5.439&
1999 CX118& 7.1&
43.662 \\
\cline{1-4} \cline{6-9} 5.180& 2002 CW154& 6.4& 46.236& & 5.440&
1998 WT31& 7.1&
45.842 \\
\cline{1-4} \cline{6-9} 5.181& 2002 CQ154& 6.4& 43.161& & 5.441&
1997 QH4& 7.1&
42.551 \\
\cline{1-4} \cline{6-9} 5.182& 2001 RW143& 6.4& 42.795& & 5.442&
2001 UO18& 7.1&
39.462 \\
\cline{1-4} \cline{6-9} 5.183& 2001 PK47& 6.4& 39.576& & 5.443&
1996 TQ66& 7.1&
39.252 \\
\cline{1-4} \cline{6-9} 5.184& 2001 KN76& 6.4& 44.022& & 5.444&
1999 RY215& 7.1&
45.128 \\
\cline{1-4} \cline{6-9} 5.185& 2001 KL76& 6.4& 44.711& & 5.445&
1994 TB& 7.1&
39.278 \\
\cline{1-4} \cline{6-9} 5.186& 2001 HY65& 6.4& 43.456& & 5.446&
2005 GZ206& 7.2&
44.747 \\
\cline{1-4} \cline{6-9} 5.187& 2001 DB106& 6.4& 43.314& & 5.447&
2005 GX186& 7.2&
43.493 \\
\cline{1-4} \cline{6-9} 5.188& 2000 WK183& 6.4& 44.258& & 5.448&
2005 ED300& 7.2&
43.923 \\
\cline{1-4} \cline{6-9} 5.189& 2000 CL105& 6.4& 43.154& & 5.449&
2005 EP296& 7.2&
46.832 \\
\cline{1-4} \cline{6-9} 5.190& 1996 TS66& 6.4& 43.719& & 5.450&
2004 DM71& 7.2&
43.329 \\
\cline{1-4} \cline{6-9} 5.191& 2005 PS21& 6.5& 44.916& & 5.451&
2004 DJ64& 7.2&
44.832 \\
\cline{1-4} \cline{6-9} 5.192& 2005 PP21& 6.5& 45.502& & 5.452&
2003 UK292& 7.2&
43.211 \\
\cline{1-4} \cline{6-9} 5.193& 2005 EO296& 6.5& 45.107& & 5.453&
2003 UY291& 7.2&
49.016 \\
\cline{1-4} \cline{6-9} 5.194& 2005 EE296& 6.5& 45.847& & 5.454&
2003 QN91& 7.2&
42.449 \\
\cline{1-4} \cline{6-9} 5.195& 2004 DM64& 6.5& 44.384& & 5.455&
2003 QF91& 7.2&
42.849 \\
\cline{1-4} \cline{6-9} 5.196& 2003 UZ291& 6.5& 44.951& & 5.456&
2003 QE91& 7.2&
44.313 \\
\cline{1-4} \cline{6-9} 5.197& 2003 QE112& 6.5& 42.97& & 5.457&
2003 QU90& 7.2&
43.384 \\
\cline{1-4} \cline{6-9} 5.198& 2003 LZ6& 6.5& 43.477& & 5.458&
2003 KP20& 7.2&
44.847 \\
\cline{1-4} \cline{6-9} 5.199& 2003 GF55& 6.5& 45.873& & 5.459&
2003 FK127& 7.2&
42.833 \\
\cline{1-4} \cline{6-9} 5.200& 2003 FM129& 6.5& 45.237& & 5.460&
2002 GV32& 7.2&
39.776 \\
\cline{1-4} \cline{6-9} 5.201& 2002 VE131& 6.5& 45.749& & 5.461&
2002 CA225& 7.2&
45.714 \\
\cline{1-4} \cline{6-9} 5.202& 2002 PV170& 6.5& 42.576& & 5.462&
2002 CT154& 7.2&
47.082 \\
\cline{1-4} \cline{6-9} 5.203& 2002 PE153& 6.5& 44.039& & 5.463&
2002 CS154& 7.2&
43.374 \\
\cline{1-4} \cline{6-9} 5.204& 2002 PO149& 6.5& 43.933& & 5.464&
2002 CR46& 7.2&
38.162 \\
\cline{1-4} \cline{6-9} 5.205& 2002 GY31& 6.5& 43.492& & 5.465&
2001 SE291& 7.2&
45.182 \\
\cline{1-4} \cline{6-9} 5.206& 2002 CC249& 6.5& 47.386& & 5.466&
2001 KE77& 7.2&
45.726 \\
\cline{1-4} \cline{6-9} 5.207& 2002 CY154& 6.5& 44.494& & 5.467&
2001 FE193& 7.2&
47.071 \\
\cline{1-4} \cline{6-9} 5.208& 2002 CO154& 6.5& 42.724& & 5.468&
2000 QL252& 7.2&
39.659 \\
\cline{1-4} \cline{6-9} 5.209& 2001 QG331& 6.5& 44.754& & 5.469&
2000 PY29& 7.2&
44.171 \\
\cline{1-4} \cline{6-9} 5.210& 2001 QP297& 6.5& 45.086& & 5.470&
2000 JF81& 7.2&
46.418 \\
\cline{1-4} \cline{6-9} 5.211& 2001 OU108& 6.5& 46.754& & 5.471&
2000 FW53& 7.2&
47.2 \\
\cline{1-4} \cline{6-9} 5.212& 2001 OQ108& 6.5& 45.826& & 5.472&
2000 CN114& 7.2&
44.068 \\
\cline{1-4} \cline{6-9}
\end{tabular}
\end{table}

\begin{table}[htbp]
\begin{tabular}
{|p{22pt}|p{70pt}|p{14pt}|p{30pt}|p{0pt}|p{22pt}|p{70pt}|p{14pt}|p{30pt}|}
\cline{1-4} \cline{6-9} N& Name& $H$& $a$, a.u.& ~& N& Name& $H$&
$a$, a.u. \\
\cline{1-4} \cline{6-9} 5.213& 2001 HA59& 6.5& 44.817& & 5.473&
1999 RA216& 7.2&
43.761 \\
\cline{1-4} \cline{6-9} 5.214& 2000 CG105& 6.5& 46.462& & 5.474&
1999 HG12& 7.2&
43.915 \\
\cline{1-4} \cline{6-9} 5.215& 1998 WX24& 6.5& 43.167& & 5.475&
1999 HR11& 7.2&
43.963 \\
\cline{1-4} \cline{6-9} 5.216& 1997 CW29& 6.5& 39.375& & 5.476&
1999 CU153& 7.2&
44.29 \\
\cline{1-4} \cline{6-9} 5.217& 1995 KJ1& 6.5& 43.468& & 5.477&
1999 CC119& 7.2&
45.032 \\
\cline{1-4} \cline{6-9} 5.218& 1998 VG44& 6.5& 39.148& & 5.478&
1998 BU48& 7.2&
33.354 \\
\cline{1-4} \cline{6-9} 5.219& 2005 PM21& 6.6& 43.678& & 5.479&
1995 WY2& 7.2&
46.412 \\
\cline{1-4} \cline{6-9} 5.220& 2005 EZ296& 6.6& 39.501& & 5.480&
1995 DC2& 7.2&
44.337 \\
\cline{1-4} \cline{6-9} 5.221& 2004 VU75& 6.6& 44.311& & 5.481&
1994 EV3& 7.2&
43.167 \\
\cline{1-4} \cline{6-9} 5.222& 2004 UD10& 6.6& 43.984& & 5.482&
1998 WA25& 7.2&
42.357 \\
\cline{1-4} \cline{6-9} 5.223& 2004 TV357& 6.6& 47.033& & 5.483&
1997 CR29& 7.2&
47.146 \\
\cline{1-4} \cline{6-9} 5.224& 2004 PY111& 6.6& 44.335& & 5.484&
1992 QB1& 7.2&
43.74 \\
\cline{1-4} \cline{6-9} 5.225& 2004 FU148& 6.6& 39.876& & 5.485&
2005 EZ300& 7.3&
37.341 \\
\cline{1-4} \cline{6-9} 5.226& 2003 UX292& 6.6& 44.063& & 5.486&
2004 SC60& 7.3&
39.46 \\
\cline{1-4} \cline{6-9} 5.227& 2003 UD292& 6.6& 45.998& & 5.487&
2004 ET95& 7.3&
43.785 \\
\cline{1-4} \cline{6-9} 5.228& 2003 QF113& 6.6& 43.8& & 5.488&
2004 DG77& 7.3&
43.955 \\
\cline{1-4} \cline{6-9} 5.229& 2003 QY111& 6.6& 43.164& & 5.489&
2004 DK71& 7.3&
44.319 \\
\cline{1-4} \cline{6-9} 5.230& 2003 QO91& 6.6& 44.289& & 5.490&
2003 UL292& 7.3&
44.902 \\
\cline{1-4} \cline{6-9} 5.231& 2002 VD131& 6.6& 44.907& & 5.491&
2003 UF292& 7.3&
44.783 \\
\cline{1-4} \cline{6-9} 5.232& 2002 PD155& 6.6& 43.119& & 5.492&
2002 VZ130& 7.3&
45.806 \\
\cline{1-4} \cline{6-9} 5.233& 2002 PH149& 6.6& 42.908& & 5.493&
2002 PB171& 7.3&
43.573 \\
\cline{1-4} \cline{6-9} 5.234& 2002 PF149& 6.6& 42.892& & 5.494&
2002 PP153& 7.3&
45.161 \\
\cline{1-4} \cline{6-9} 5.235& 2002 PE149& 6.6& 43.046& & 5.495&
2002 GC32& 7.3&
43.473 \\
\cline{1-4} \cline{6-9} 5.236& 2002 FW36& 6.6& 43.115& & 5.496&
2002 FU6& 7.3&
46.077 \\
\cline{1-4} \cline{6-9} 5.237& 2002 CZ224& 6.6& 45.115& & 5.497&
2002 CD251& 7.3&
42.744 \\
\cline{1-4} \cline{6-9} 5.238& 2002 CU154& 6.6& 43.954& & 5.498&
2001 KW76& 7.3&
46.304 \\
\cline{1-4} \cline{6-9} 5.239& 2001 UN18& 6.6& 44.074& & 5.499&
2001 FN185& 7.3&
42.804 \\
\cline{1-4} \cline{6-9} 5.240& 2001 UA17& 6.6& 42.915& & 5.500&
2001 DM108& 7.3&
45.437 \\
\cline{1-4} \cline{6-9} 5.241& 2001 QT322& 6.6& 36.953& & 5.501&
2000 YY142& 7.3&
44.9 \\
\cline{1-4} \cline{6-9} 5.242& 2001 QA298& 6.6& 46.186& & 5.502&
2000 WM183& 7.3&
44.851 \\
\cline{1-4} \cline{6-9} 5.243& 2001 KH76& 6.6& 46.351& & 5.503&
2000 WL183& 7.3&
43.138 \\
\cline{1-4} \cline{6-9} 5.244& 2000 QC226& 6.6& 43.993& & 5.504&
2000 QN251& 7.3&
41.984 \\
\cline{1-4} \cline{6-9} 5.245& 2000 QB226& 6.6& 45.644& & 5.505&
2000 QD226& 7.3&
41.171 \\
\cline{1-4} \cline{6-9} 5.246& 2000 OL67& 6.6& 44.986& & 5.506&
2000 FH8& 7.3&
43.779 \\
\cline{1-4} \cline{6-9} 5.247& 2000 GP183& 6.6& 40.083& & 5.507&
2000 CS105& 7.3&
44.713 \\
\cline{1-4} \cline{6-9} 5.248& 2000 CQ114& 6.6& 46.298& & 5.508&
1999 SA28& 7.3&
39.234 \\
\cline{1-4} \cline{6-9} 5.249& 2000 CY105& 6.6& 45.533& & 5.509&
1999 RC216& 7.3&
44.403 \\
\cline{1-4} \cline{6-9} 5.250& 1999 HS11& 6.6& 44.413& & 5.510&
1999 RB216& 7.3&
47.316 \\
\cline{1-4} \cline{6-9} 5.251& 1998 WY31& 6.6& 45.136& & 5.511&
1999 RU215& 7.3&
42.866 \\
\cline{1-4} \cline{6-9} 5.252& 1998 WX31& 6.6& 45.318& & 5.512&
1999 RW214& 7.3&
42.727 \\
\cline{1-4} \cline{6-9} 5.253& 1998 WG24& 6.6& 45.513& & 5.513&
1999 OD4& 7.3&
41.403 \\
\cline{1-4} \cline{6-9} 5.254& 1998 KG62& 6.6& 43.389& & 5.514&
1999 CH119& 7.3&
43.401 \\
\cline{1-4} \cline{6-9} 5.255& 1997 CT29& 6.6& 43.891& & 5.515&
1999 CD119& 7.3&
43.993 \\
\cline{1-4} \cline{6-9} 5.256& 2001 KP76& 6.6& 43.772& & 5.516&
1998 KY61& 7.3&
44.326 \\
\cline{1-4} \cline{6-9} 5.257& 2000 FD8& 6.6& 44.035& & 5.517&
1997 RT5& 7.3&
41.166 \\
\cline{1-4} \cline{6-9} 5.258& 1997 CQ29& 6.6& 45.423& & 5.518&
1996 KV1& 7.3&
45.413 \\
\cline{1-4} \cline{6-9} 5.259& 1999 HU11 $^{10}$ & 6.6& 44.382& &
5.519& 2000 PK30& 7.3&
38.625 \\
\cline{1-4} \cline{6-9} 5.260& 1997 CU29& 6.6& 43.517& & 5.520&
1999 RA215& 7.3&
43.078 \\
\cline{1-4} \cline{6-9}
\end{tabular}
\end{table}

\begin{table}[htbp]
\caption{Near region of scattered-disk Kuiper belt}
\begin{tabular}
{|p{20pt}|p{68pt}|p{16pt}|p{30pt}|p{0pt}|p{20pt}|p{68pt}|p{14pt}|p{30pt}|}
\cline{1-4} \cline{6-9} N& Name& $H$& $a$, a.u.& ~& N& Name& $H$&
$a$, a.u. \\
\cline{1-4} \cline{6-9} 6.1& 2003 UB313& $-1.2$& 67.668& & 6.17&
2002 CY224& 6.3&
54.085 \\
\cline{1-4} \cline{6-9} 6.2& 2002 TC302& 3.9& 55.037& & 6.18& 2004
OJ14& 6.4&
55.085 \\
\cline{1-4} \cline{6-9} 6.3& 2001 UR163& 4.2& 51.357& & 6.19& 2002
GZ31& 6.5&
50.845 \\
\cline{1-4} \cline{6-9} 6.4& 1999 DE9& 4.7& 55.856& & 6.20& 1999
HW11& 6.5&
53.079 \\
\cline{1-4} \cline{6-9} 6.5& 2000 YW134& 5.0& 57.896& & 6.21& 2002
GP32& 6.7&
56.063 \\
\cline{1-4} \cline{6-9} 6.6& 1995 TL8& 5.4& 52.239& & 6.22& 2001
KC77& 6.7&
55.642 \\
\cline{1-4} \cline{6-9} 6.7& 2002 JR146& 5.5& 53.398& & 6.23& 2000
FE8& 6.7&
55.884 \\
\cline{1-4} \cline{6-9} 6.8& 2001 QX322& 5.7& 59.959& & 6.24& 2000
CM114& 6.8&
59.938 \\
\cline{1-4} \cline{6-9} 6.9& 2001 QW297& 5.7& 51.594& & 6.25& 2003
QK91& 6.9&
67.574 \\
\cline{1-4} \cline{6-9} 6.10& 1999 CC158& 5.8& 54.024& & 6.26&
2002 GA32& 7.0&
52.423 \\
\cline{1-4} \cline{6-9} 6.11& 2000 CQ105& 5.9& 57.345& & 6.27&
2002 GG32& 7.1&
55.884 \\
\cline{1-4} \cline{6-9} 6.12& 2004 PD112& 6.1& 64.295& & 6.28&
2002 CX154& 7.1&
71.934 \\
\cline{1-4} \cline{6-9} 6.13& 2001 KG76& 6.2& 51.996& & 6.29& 2000
YY1& 7.1&
63.243 \\
\cline{1-4} \cline{6-9} 6.14& 2000 PE30& 6.2& 54.471& & 6.30& 1999
HB12& 7.2&
56.296 \\
\cline{1-4} \cline{6-9} 6.15& 1998 XY95& 6.2& 64.215& & 6.31& 2002
GX32& 7.3&
53.671 \\
\cline{1-4} \cline{6-9} 6.16& 2003 UY117& 6.3& 55.19& & & & &
 \\
\cline{1-4} \cline{6-9}
\end{tabular}
\label{tab6}
\end{table}

\begin{table}[htbp]
\caption{Distant region of scattered-disk Kuiper belt}
\begin{tabular}
{|p{22pt}|p{70pt}|p{14pt}|p{30pt}|p{0pt}|p{22pt}|p{70pt}|p{14pt}|p{30pt}|}
\cline{1-4} \cline{6-9} N& Name& $H$& $a$, a.u.& ~& N& Name& $H$&
$a$, a.u. \\
\cline{1-4} \cline{6-9} 7.1& 2003 VB12 $^{12}$ & 1.6& 489& & 7.7&
2001 FZ173& 6.2&
87.049 \\
\cline{1-4} \cline{6-9} 7.2& 1996 GQ21& 5.2& 95.067& & 7.8& 2003
FX128& 6.3&
103 \\
\cline{1-4} \cline{6-9} 7.3& 1996 TL66& 5.4& 82.781& & 7.9& 2000
OM67& 6.7&
97.941 \\
\cline{1-4} \cline{6-9} 7.4& 2004 TF282& 6.0& 80.044& & 7.10& 2004
PB112& 7.2&
107 \\
\cline{1-4} \cline{6-9} 7.5& 2000 CR105& 6.1& 221& & 7.11& 1999
RU214& 7.3&
95.528 \\
\cline{1-4} \cline{6-9} 7.6& 2001 FP185& 6.1& 227& & 7.12& 1999
CF119& 7.3&
89.602 \\
\cline{1-4} \cline{6-9}
\end{tabular}
\label{tab7}
\end{table}

\begin{table}[htbp]
\caption{Exoplanets}
\begin{tabular}
{|p{17pt}|p{67pt}|p{27pt}|p{30pt}|p{0pt}|p{19pt}|p{67pt}|p{25pt}|p{30pt}|}
\cline{1-4} \cline{6-9} N& Name& $m/m_{J}$& $a$, a.u.& ~& N& Name&
$m/m_{J}$&
$a$, a.u. \\
\cline{1-4} \cline{6-9} 8.1& PSR B1257+12 b$^{13}$&
$6.3\times10^{-5}$& 0.19& & 8.95& HD 4203~b& 1.65&
1.09 \\
\cline{1-4} \cline{6-9} 8.2& PSR B1257+12 d$^{14}$& 0.0123& 0.46&
& 8.96& HD 160691~b& 1.67&
1.5 \\
\cline{1-4} \cline{6-9} 8.3& PSR B1257+12 c$^{15}$& 0.0135& 0.36&
& 8.97& 16 Cyg B~b& 1.69&
1.67 \\
\cline{1-4} \cline{6-9} 8.4& Gl 876~d& 0.023& 0.021& & 8.98& HD
50499~b& 1.71&
3.86 \\
\cline{1-4} \cline{6-9} 8.5& HD 160691~d& 0.044& 0.09& & 8.99& HD
82943~b& 1.75&
$\sim 1.19$ \\
\cline{1-4} \cline{6-9} 8.6& HD 212301~b& 0.045& 0.036& & 8.100&
HD 154857~b& 1.8&
1.11 \\
\cline{1-4} \cline{6-9} 8.7& 55 Cnc~e& 0.045& 0.038& & 8.101& HD
196885~b& 1.84&
1.12 \\
\cline{1-4} \cline{6-9} 8.8& HD 4308~b& 0.047& 0.114& & 8.102& HD
74156~b& 1.86&
0.294 \\
\cline{1-4} \cline{6-9} 8.9& Gl 581~b& 0.056& 0.041& & 8.103& HD
73256~b& 1.87&
0.037 \\
\cline{1-4} \cline{6-9} 8.10& HD 190360~c& 0.057& 0.128& & 8.104&
Proxima $^{ 20}$& 1.89&
1.3 \\
\cline{1-4} \cline{6-9} 8.11& GJ 436~b& 0.067& 0.028& & 8.105&
$\nu $ And c& 1.89&
0.829 \\
\cline{1-4} \cline{6-9} 8.12& HD 49674~b& 0.11& 0.057& & 8.106& HD
68988~b& 1.9&
0.071 \\
\cline{1-4} \cline{6-9} 8.13& HD 11964~b& 0.11& 0.229& & 8.107& Gl
876~b& 1.94&
0.208 \\
\cline{1-4} \cline{6-9}
\end{tabular}
\end{table}

\begin{table}[htbp]
\begin{tabular}
{|p{17pt}|p{75pt}|p{27pt}|p{30pt}|p{0pt}|p{19pt}|p{75pt}|p{25pt}|p{30pt}|}
\cline{1-4} \cline{6-9} N& Name& $m/m_{J}$& $a$, a.u.& ~& N& Name&
$m/m_{J}$&
$a$, a.u. \\
\cline{1-4} \cline{6-9} 8.14& HD 99492~b& 0.122& 0.119& & 8.108&
HR 810~b& 1.94&
0.91 \\
\cline{1-4} \cline{6-9} 8.15& HD 102117~b& 0.14& 0.149& & 8.109&
HD 19994~b& 2.0&
1.3 \\
\cline{1-4} \cline{6-9} 8.16& HD 117618~b& 0.19& 0.28& & 8.110& HD
70642~b& 2.0&
3.3 \\
\cline{1-4} \cline{6-9} 8.17& HD 76700~b& 0.197& 0.049& & 8.111&
OGLE 235-MOA53 b$^{ 21}$& 2&
2.9 d \\
\cline{1-4} \cline{6-9} 8.18& HD 3651~b& 0.2& 0.284& & 8.112& HD
82943~c& 2.01&
$\sim 0.746$ \\
\cline{1-4} \cline{6-9} 8.19& 55 Cnc~c& 0.217& 0.24& & 8.113& HD
117207~b& 2.06&
3.78 \\
\cline{1-4} \cline{6-9} 8.20& HD 88133~b& 0.22& 0.047& & 8.114& HD
217107~c& 2.1&
4.3 \\
\cline{1-4} \cline{6-9} 8.21& HD 168746~b& 0.23& 0.065& & 8.115&
HD 216437~b& 2.1&
2.7 \\
\cline{1-4} \cline{6-9} 8.22& HD 16141~b& 0.23& 0.35& & 8.116& HD
118203~b& 2.13&
0.07 \\
\cline{1-4} \cline{6-9} 8.23& HD 46375~b& 0.249& 0.041& & 8.117&
HD 128311~b& 2.18&
1.099 \\
\cline{1-4} \cline{6-9} 8.24& HD 109749~b& 0.28& 0.064& & 8.118&
HD 8574~b& 2.23&
0.76 \\
\cline{1-4} \cline{6-9} 8.25& HD 101930~b& 0.3& 0.302& & 8.119& HD
37605~b& 2.3&
0.25 \\
\cline{1-4} \cline{6-9} 8.26& PSR B1257+12 e $^{16}$& $\sim 0.3$&
40& & 8.120& HD 12661~b& 2.3&
0.83 \\
\cline{1-4} \cline{6-9} 8.27& HD 149026~b& 0.36& 0.042& & 8.121&
HD 41004 A~b& 2.3&
1.31 \\
\cline{1-4} \cline{6-9} 8.28& HD 93083~b& 0.37& 0.477& & 8.122& HD
202206~c& 2.44&
2.55 \\
\cline{1-4} \cline{6-9} 8.29& HD 63454~b& 0.38& 0.036& & 8.123& 47
UMa~b& 2.54&
2.09 \\
\cline{1-4} \cline{6-9} 8.30& HD 108147~b& 0.4& 0.104& & 8.124& HD
23079~b& 2.61&
1.65 \\
\cline{1-4} \cline{6-9} 8.31& HD 83443~b& 0.41& 0.04& & 8.125&
OGLE-05-071 b $^{22}$& 2.7&
3 d \\
\cline{1-4} \cline{6-9} 8.32& HD 75289~b& 0.42& 0.046& & 8.126& HD
89307~b& 2.73&
4.15 \\
\cline{1-4} \cline{6-9} 8.33& HD 208487~b& 0.45& 0.49& & 8.127& HD
169830~b& 2.88&
0.81 \\
\cline{1-4} \cline{6-9} 8.34& 51 Peg~b& 0.468& 0.052& & 8.128& HD
219449~b& 2.9&
0.3 \\
\cline{1-4} \cline{6-9} 8.35& HD 2638~b& 0.48& 0.044& & 8.129& HD
72659~b& 2.96&
4.16 \\
\cline{1-4} \cline{6-9} 8.36& BD-10 3166~b& 0.48& 0.046& & 8.130&
HD 73526~b& 3.0&
0.66 \\
\cline{1-4} \cline{6-9} 8.37& HD 6434~b& 0.48& 0.15& & 8.131& HD
196050~b& 3.0&
2.5 \\
\cline{1-4} \cline{6-9} 8.38& HD 102195~b& 0.48& 0.049& & 8.132&
HD 160691~c& 3.1&
4.17 \\
\cline{1-4} \cline{6-9} 8.39& HD 187123~b& 0.52& 0.042& & 8.133&
61 Cyg B b $^{ 23}$& 3.14&
3.3 \\
\cline{1-4} \cline{6-9} 8.40& OGLE-TR-111~b& 0.53& 0.047& & 8.134&
HD 128311~c& 3.21&
1.76 \\
\cline{1-4} \cline{6-9} 8.41& OGLE-TR-10~b& 0.54& 0.042& & 8.135&
GJ 3021~b& 3.32&
0.49 \\
\cline{1-4} \cline{6-9} 8.42& Gl 876~c& 0.56& 0.13& & 8.136& HD
40979~b& 3.32&
0.811 \\
\cline{1-4} \cline{6-9} 8.43& HD 34445~b& 0.58& 0.51& & 8.137& HD
80606~b& 3.41&
0.439 \\
\cline{1-4} \cline{6-9} 8.44& HD 37124~c& 0.6& 1.64& & 8.138& HD
195019~b& 3.43&
0.14 \\
\cline{1-4} \cline{6-9} 8.45& TrES-1~& 0.61& 0.039& & 8.139& 61
Cyg A b $^{ 24}$& 3.67&
2.9 \\
\cline{1-4} \cline{6-9} 8.46& HD 37124~b& 0.61& 0.53& & 8.140& HD
183263~b& 3.69&
1.52 \\
\cline{1-4} \cline{6-9} 8.47& HD 27894~b& 0.62& 0.122& & 8.141&
$\nu $ And~d& 3.75&
2.53 \\
\cline{1-4} \cline{6-9} 8.48& HD 216770~b& 0.65& 0.46& & 8.142& HD
92788~b& 3.86&
0.97 \\
\cline{1-4} \cline{6-9} 8.49& HD 37124~d& 0.66& 3.19& & 8.143&
$\tau$ Boo~b& 3.9&
0.046 \\
\cline{1-4} \cline{6-9} 8.50& HD 209458~b $^{17}$& 0.69& 0.045& &
8.144& 55 Cnc~d& 3.92&
5.257 \\
\cline{1-4} \cline{6-9} 8.51& $\nu$ And~b& 0.69& 0.059& & 8.145&
Gl 86~b& 4.01&
0.11 \\
\cline{1-4} \cline{6-9} 8.52& HD 192263~b& 0.72& 0.15& & 8.146& HD
169830~c& 4.04&
3.6 \\
\cline{1-4} \cline{6-9} 8.53& HD 330075~b& 0.76& 0.043& & 8.147&
HD 142022 A~b& 4.4&
2.8 \\
\cline{1-4} \cline{6-9} 8.54& 47 UMa~c& 0.76& 3.73& & 8.148& HD
213240~b& 4.5&
2.03 \\
\cline{1-4} \cline{6-9} 8.55& HD 38529~b& 0.78& 0.129& & 8.149& 14
Her~b& 4.74&
2.8 \\
\cline{1-4} \cline{6-9}
\end{tabular}
\end{table}

\begin{table}[htbp]
\begin{tabular}
{|p{17pt}|p{75pt}|p{27pt}|p{30pt}|p{0pt}|p{19pt}|p{75pt}|p{25pt}|p{30pt}|}
\cline{1-4} \cline{6-9} N& Name& $m/m_{J}$& $a$, a.u.& ~& N& Name&
$m/m_{J}$&
$a$, a.u. \\
\cline{1-4} \cline{6-9} 8.56& 55 Cnc~b& 0.784& 0.115& & 8.150& HD
2039~b& 4.85&
2.19 \\
\cline{1-4} \cline{6-9} 8.57& HD 4208~b& 0.8& 1.67& & 8.151& HD
50554~b& 4.9&
2.38 \\
\cline{1-4} \cline{6-9} 8.58& HD 114729~b& 0.82& 2.08& & 8.152& HD
47536~b & 4.96&
1.61 \\
\cline{1-4} \cline{6-9} 8.59& $\varepsilon$ Eri~b& 0.86& 3.3& &
8.153& HD 190228~b& 4.99&
2.31 \\
\cline{1-4} \cline{6-9} 8.60& HD 121504~b& 0.89& 0.32& & 8.154&
PSR B1620-26 b $^{25}$& $\sim 5$&
10--30 \\
\cline{1-4} \cline{6-9} 8.61& Lalande 21185 b $^{18}$& 0.9& 2.4& &
8.155& 2M 1207 b $^{ 26}$& 5&
41 d \\
\cline{1-4} \cline{6-9} 8.62& HD 10647~b& 0.91& 2.1& & 8.156& HD
222582~b& 5.11&
1.35 \\
\cline{1-4} \cline{6-9} 8.63& HD 179949~b& 0.98& 0.04& & 8.157& HD
59686~b& 5.25&
0.911 \\
\cline{1-4} \cline{6-9} 8.64& HD 45350~b& 0.98& 1.77& & 8.158& HD
28185~b& 5.7&
1.03 \\
\cline{1-4} \cline{6-9} 8.65& HD 114783~b& 0.99& 1.2& & 8.159& 61
Cyg A c $^{27}$& 5.76&
4.7 \\
\cline{1-4} \cline{6-9} 8.66& HD 114386~b& 0.99& 1.62& & 8.160&
DM+59$^{o}$1915 $^{28}$& 6&
 \\
\cline{1-4} \cline{6-9} 8.67& HD 150706~b& 1.0& 0.82& & 8.161& HD
10697~b& 6.12&
2.13 \\
\cline{1-4} \cline{6-9} 8.68& HD 142~b& 1.0& 0.98& & 8.162& HD
74156~c& 6.17&
3.4 \\
\cline{1-4} \cline{6-9} 8.69& HD 147513~b& 1.0& 1.26& & 8.163& HD
178911 B~b& 6.29&
0.32 \\
\cline{1-4} \cline{6-9} 8.70& HD 108874~c& 1.018& 2.68& & 8.164&
HD 104985~b& 6.3&
0.78 \\
\cline{1-4} \cline{6-9} 8.71& $\rho$ CrB~b& 1.04& 0.22& & 8.165&
HD 11977~b& 6.54&
1.93 \\
\cline{1-4} \cline{6-9} 8.72& HD 20367~b& 1.07& 1.25& & 8.166& HD
111232~b& 6.8&
1.97 \\
\cline{1-4} \cline{6-9} 8.73& HD 130322~b& 1.08& 0.088& & 8.167&
HD 106252~b& 6.81&
2.61 \\
\cline{1-4} \cline{6-9} 8.74& HD 52265~b& 1.13& 0.49& & 8.168& HD
81040~b& 6.86&
1.94 \\
\cline{1-4} \cline{6-9} 8.75& HD 188753A~b& 1.14& 0.0446& & 8.169&
HD 23596~b& 7.19&
2.72 \\
\cline{1-4} \cline{6-9} 8.76& HD 189733~b& 1.15& 0.0313& & 8.170&
HD 168443~b& 7.2&
0.29 \\
\cline{1-4} \cline{6-9} 8.77& OGLE-TR-132~b& 1.19& 0.0306& &
8.171& 70 Vir~b& 7.44&
0.48 \\
\cline{1-4} \cline{6-9} 8.78& HD 65216~b& 1.21& 1.37& & 8.172& HD
89744~b& 7.99&
0.89 \\
\cline{1-4} \cline{6-9} 8.79& HD 210277~b& 1.24& 1.097& & 8.173&
HIP 75458~b& 8.64&
1.34 \\
\cline{1-4} \cline{6-9} 8.80& HD 188015~b& 1.26& 1.19& & 8.174& EV
Lac b $^{29}$& 9--23&
$\sim 5$ \\
\cline{1-4} \cline{6-9} 8.81& HD 177830~b& 1.28& 1& & 8.175& HD
33564~b& 9.1&
1.1 \\
\cline{1-4} \cline{6-9} 8.82& HD 27442~b& 1.28& 1.18& & 8.176& HD
30177~b& 9.17&
3.86 \\
\cline{1-4} \cline{6-9} 8.83& HD 149143~b& 1.33& 0.053& & 8.177&
HD 33636~b& 9.28&
3.56 \\
\cline{1-4} \cline{6-9} 8.84& OGLE-TR-113~b& 1.35& 0.0228& &
8.178& HD 141937~b& 9.7&
1.52 \\
\cline{1-4} \cline{6-9} 8.85& HD 108874~b& 1.36& 1.051& & 8.179&
Kr\"{u} 60(A/B) b $^{ 30}$& 10&
 \\
\cline{1-4} \cline{6-9} 8.86& HD 217107~b& 1.37& 0.074& & 8.180&
HD 39091~b& 10.35&
3.29 \\
\cline{1-4} \cline{6-9} 8.87& OGLE-TR-56~b& 1.45& 0.0225& & 8.181&
HD 114762~b& 11.02&
0.3 \\
\cline{1-4} \cline{6-9} 8.88& HD 216435~b& 1.49& 2.7& & 8.182& HD
136118~b& 11.9&
2.3 \\
\cline{1-4} \cline{6-9} 8.89& HD 190360~b& 1.502& 3.92& & 8.183&
HD 38529~c& 12.7&
3.68 \\
\cline{1-4} \cline{6-9} 8.90& HD 12661~c& 1.57& 2.56& & 8.184& AB
Pic b $^{ 31}$& 13.5&
275 d \\
\cline{1-4} \cline{6-9} 8.91& HD 134987~b& 1.58& 0.78& & 8.185& HD
162020~b& 13.75&
0.072 \\
\cline{1-4} \cline{6-9} 8.92& $\gamma $ Cep~b& 1.59& 2.03& &
8.186& HD 13189~b& 14&
1.85 \\
\cline{1-4} \cline{6-9} 8.93& Lalande 21185 c $^{19}$& 1.6& 7.3& &
8.187& HD 168443~c& 17.1&
2.87 \\
\cline{1-4} \cline{6-9} 8.94& HD 142415~b& 1.62& 1.05& & 8.188& HD
202206~b& 17.4&
0.83 \\
\cline{1-4} \cline{6-9}
\end{tabular}
\label{tab8}
\end{table}

\clearpage

\begin{table}[ht]
\caption{Candidates}
\begin{tabular}
{|p{16pt}|p{70pt}|p{27pt}|p{30pt}|p{0pt}|p{16pt}|p{70pt}|p{26pt}|p{30pt}|}
\cline{1-4} \cline{6-9} N& Name& $m$/$m_{J}$& $a$, a.u.& ~& N&
Name& $m$/$m_{J}$&
$a$, a.u. \\
\cline{1-4} \cline{6-9} 9.1& 98-BLG-35$^{32}$& $\sim 0.003$& $\sim
80$& & 9.12& 95-BLG-3 $^{ 38}$& $\sim 2$&
$> 5-10$ \\
\cline{1-4} \cline{6-9} 9.2& PSR B0329+54 b$^{ 33}$& $> 0.0063$&
7.3& & 9.13& 14 Her c& 2.086&
5.81 \\
\cline{1-4} \cline{6-9} 9.3& E Eri c& 0.1& 40& & 9.14& HD 62509 b&
2.9&
1.9 \\
\cline{1-4} \cline{6-9} 9.4& Anon b& 0.3& 0.024& & 9.15& 97-BLG-41
$^{39}$& 3.5&
7 \\
\cline{1-4} \cline{6-9} 9.5& HD $\!219542$ B b& 0.3& 0.46& & 9.16&
94-BLG-4 $^{40}$& $\sim 5$&
$\sim $~1 \\
\cline{1-4} \cline{6-9} 9.6& HD 208487 c& ?$\:0.46$& ? 1.92& &
9.17& VB 8 $^{41}$& 10&
 \\
\cline{1-4} \cline{6-9} 9.7& OGLE-TR-36 $^{ 34}$& 0.5& 0.025& &
9.18& PSR B1620--26 b $^{42}$& $\sim 5$&
10--30 \\
\cline{1-4} \cline{6-9} 9.8& Barnard d$^{ 35}$& 0.63& 2& & 9.19&
HD 8673 b& 14&
1.58 \\
\cline{1-4} \cline{6-9} 9.9& HD 11964 c& 0.7& 3.167& & 9.20& HD
100546 b& $\sim 20$&
? 6.5 \\
\cline{1-4} \cline{6-9} 9.10& Barnard c $^{36}$& 0.84& 3& & 9.21&
GQ Lup b $^{43}$& 21.5&
103 d \\
\cline{1-4} \cline{6-9} 9.11& Barnard b $^{ 37}$& 1.26& 5& & &
 \quad &
&
 \\
\cline{1-4} \cline{6-9}
\end{tabular}
\label{tab9}
\end{table}

{\small1 -- Chariklo, 2 -- Chiron, 3 -- Pholus, 4 -- Orcus, 5 --
Quaoar, 6 -- Ixion, 7 -- Varuna, 8 -- Huya, 9 -- Chaos, 10 --
Deucalion, 11 -- Rhadamanthus, 12 -- Sedna; 13 -- Me, RV; 14 --
Si/Me, RV; 15 -- Si/Me, RV; 16 -- H-He/Ic, RV; 17 -- HPh+RV; 18 --
A; 19 -- A; 20 -- A; 21 -- GML; 22 -- GML; 23 -- A; 24 -- A; 25 --
RV; 26 -- IF; 27 -- A; 28 -- A; 29 -- A; 30 -- A, 31 -- IF; 32 --
GML; 33 -- Ic/H-He, RV; 34 -- GML; 35 -- A.; 36 -- A.; 37 -- A.;
38 -- GML; 39 -- GML; 40 -- GML; 41 -- IR; 42 -- H-He, RA; 43 --
IF.}

\smallskip \hrule \medskip

Thus, optimistic estimations of the number of Solar system planets
gives 600 candidates: 9 big planet + 20 satellites of the big
planets + 3 Main asteroid belt objects + 5 centaurs + 563 Kuiper
belt asteroids (with $A_{H}= 0.03$, $H=7^{m}.3$). Corresponding
pessimistic estimations are 122 planets = 9 + 20 + 3 + 0 + 90
(with $A_{H}= 0.12$, $H=5^{m}.8$). Discoveries of 188 planets are
enough reliable and 21 exoplanets are needed additional analysis.

\begin {references}

\reference Alexandrov Yu.V., Zakhozhay V.A. 1980, in Astron. vest.
(Rus), 14(3), 129.

\reference Sluta E.N., Voropaev S.A. 1980, in Astron. vest. (Rus).
27(1), 71.

\reference Zakhozhay O.V., Zakhozhay V.A., Krugly Yu.N. 2006a, in
Abstracts of 13$^{th}$ Open Young Scientists' Conference on
Astronomy and Space Physics 25-29 April 2006, Kyiv National Taras
Shevchenko University, 110.

\reference Zakhozhay O.V., Krugly Yu.N., Zakhozhay V.A. 2006b, in
Abstracts of student scientific conf. physics school 18 April
2006, Kharkov: V.N.Karazin Kharkiv National University, 22.

\reference Zakhozhay V.A. 2001, in Extension and connection of
reference frames using ground based CCD technique, Nikolaev, 274.

\reference Zakhozhay V.A. 2005a, in Visn. astron. shkoly (Ukr),
4(2), 34.

\reference Zakhozhay V.A. 2005b, in Visn. astron. shkoly (Ukr),
4(2), 55.

\end {references}

\end{document}